\documentclass[11pt]{article}
\usepackage{fullpage}
\usepackage{booktabs}
\usepackage{makecell}
\usepackage{pdfpages}

\usepackage[utf8]{inputenc} 
\usepackage[T1]{fontenc}    
\usepackage{hyperref, bm, lmodern}       
\usepackage{url}            
\usepackage{booktabs}       
\usepackage{amsmath,amsfonts,mathtools}       
\usepackage{caption,enumerate} 

\usepackage{algorithm} 
\usepackage{algpseudocode}

\usepackage{float,amssymb,mathtools,mathrsfs,placeins,amsthm}
\usepackage{nicefrac}       

\usepackage{lipsum,setspace}
\usepackage{graphicx,xcolor,comment}
\usepackage{natbib} 
\usepackage{subcaption}

\newtheorem{proposition}{Proposition}

\theoremstyle{remark}
\newtheorem{remark}{Remark}
\newtheorem{definition}{Definition}

\usepackage{mathabx}
\usepackage{cleveref}

\allowdisplaybreaks 

\newcommand{\bmu}{\bm{\mu}}

\usepackage{setspace}
\title{Bayesian Bradley-Terry Framework for Ranking Multi-Indicator Entities: Application to Indian States and Union Territories}

\author{
Arshi Rizvi\\
Department of Mathematics\\
Indian Institute of Technology Delhi\\
Hauz Khas, Delhi 110016, India\\
\texttt{maz248174@maths.iitd.ac.in}
\and
Rahul Singh\\
Department of Mathematics\\
Indian Institute of Technology Delhi\\
Hauz Khas, Delhi 110016, India\\
\texttt{sirahul@iitd.ac.in}
}

\date{}

\begin{document}
\maketitle

\begin{abstract}
Often, real-world entities are explained by multiple indicators, and ranking those is challenging, e.g.,  administrative units, sport teams, and machine learning models. Existing methodologies rely on single indicators or composite indices, leaving a gap for frameworks capable of handling multi-indicator scenarios. We present a ranking methodology motivated by the Bayesian Bradley-Terry (BT) model. We convert indicators to paired comparisons to utilize the BT framework. The prior covariance of the BT latent merit parameters is modeled as a function of a covariate to accommodate prior information. For parameter estimation, we use a Markov chain Monte Carlo (MCMC) algorithm that employs a hybrid Metropolis--Hastings scheme with a preconditioned Crank--Nicolson proposal and Gibbs sampling. We show that the proposed estimator converges. Furthermore, we propose a problem-specific stopping rule based on ranking stability. To demonstrate the methodology's practical utility, we apply it to rank the states and union territories (UTs) of India using multi-indicator data from the National Family Health Survey-5. By estimating and comparing rankings across various economic regimes, including low-income, mid-income, and outlier-excluded subsets, our application uncovers meaningful deviations between regional economic standing and overall developmental performance, illustrating the framework's broad value for resolving complex multidimensional ranking problems.
\end{abstract}


\section{Introduction}
Ranking complex entities, such as administrative units, machine learning models, or socio-economic units, based on multiple performance indicators is a critical challenge in applied statistics and evidence-based decision-making. Most existing methodologies are not designed to handle high-dimensional data and to provide an explainable ranking. Index-based approaches, such as the Human Development Index, aggregate indicators into a single score but struggle when individual metrics possess relatively small magnitudes \citep{klugman2011hdi}. For high-dimensional data, Principal Component Analysis (PCA) is frequently employed; however, it often yields arbitrary, difficult-to-interpret results that lack direct policy relevance \citep{vyas2006constructing}. Another approach involves linear combinations of indicator variables, with coefficients estimated via expert-generated pairwise comparisons (\citealt{floridi2022pairwise}); this is restricted to only a handful of indicators. Therefore, there is a need for a rigorous statistical methodology capable of handling a large number of indicators to meaningfully address real--world ranking problems.

To address this methodological gap, we present an inclusive and adaptable ranking framework inspired by paired-comparisons methods. By evaluating objects in pairs based on distinct properties, paired comparisons have been successfully applied to rank diverse items, from sports teams \citep{beaudoin2018computationally, bozoki2016application} and chess players \citep{csato2017ranking} to Large Language Model chatbots \citep{ameli2025a}. A popular model to obtain rankings from these paired comparisons is the Bradley-Terry (BT) model (\citealt{bradley1952rank}). A Bayesian BT framework enables inclusion of relevant prior knowledge. This adaptability is important in applied settings; researchers can incorporate external covariates, such as baseline economic characteristics, spatial proximity, or historical performance, to stabilize and refine merit estimation. The Bayesian BT model has been studied within known paired-comparison frameworks, e.g., comparisons of machine learning algorithms \citep{wainer2023bayesian} and the identification of deprived urban regions \citep{seymour2022bayesian}. However, it is not known how to incorporate multiple indicator scenarios in the BT framework to the best of our knowledge.

We propose a Bayesian BT model for multi-indicator scenarios. To estimate the merit parameters from the posterior distribution, we implement an efficient Markov chain Monte Carlo (MCMC) sampling scheme. By combining a preconditioned Crank-Nicolson (pCN) proposal with Gibbs sampling, the methodology resolves the non-conjugacy of the prior. We establish convergence of this MCMC approach and introduce a novel, problem-specific stopping rule based on ranking stability, thereby providing a comprehensive, end-to-end framework for applied researchers. 

To demonstrate the methodology's practical utility and its capacity to analyze high-dimensional survey data, we apply the framework to evaluate the developmental progress of India's 36 administrative units (28 states and 8 union territories). Despite decentralized governance, substantial heterogeneity persists across India in outcomes such as literacy, health, and income \citep{nayyar2008economic, bajpai1996trends, lolayekar2017growth}. We use data from the National Family Health Survey-5 (NFHS-5) to construct pairwise comparisons of family well-being indicators, including disease prevalence, education, and living standards. Recognizing that economic capacity is a fundamental driver of societal development \citep{thorn1968per, hordofa2023moderating}, we model the prior covariance of the BT merit parameters as a function of per capita income. While previous studies have analyzed NFHS data across specific, isolated dimensions \citep{gopika2022status, vignitha2024women}, our framework provides a unified, data-driven ranking of these regions. Finally, this application not only reveals critical disparities between regional economic standing and overall developmental performance but also illustrates how this generalized methodology can facilitate evidence-based policymaking in complex multidimensional settings.

The rest of the paper is organised as follows. The BT, Bayesian BT models and datasets used are introduced in Section \ref{section:prelim}. Section \ref{methodology_section} discusses the ranking methodology, the prior and modelling of prior's covariance, followed by parameter estimation with its diagnostics and stopping rule. Section \ref{NFHS_implementation} presents the implementation of the methodology on the NFHS-5 data, yielding rankings for different subsets of states/UTs that are then compared. We conclude in Section \ref{discussion} with a summary and discussion of possible extensions of our methodology.

\section{Preliminaries}
\label{section:prelim}
In this section, we describe the Bradley-Terry (BT) model, the Bayesian BT model, and the datasets used. 
\subsection{The BT Model}
\label{BTM}
The BT model was first introduced by \cite{zermelo1929berechnung} and later independently proposed by \cite{bradley1952rank}.
To formalize the model, consider a set of $M$ items with merit parameter vector $\bmu= (\mu_1,\mu_2,\ldots,\mu_M)^\top$, where $\mu_i$ is merit parameter of $i^{th}$ item. Suppose there are $K$ indicators, based on which the performance of $M$ items is to be analyzed. Each pair of items is compared $K$ times, so the total number of comparisons is $KM(M-1)/2$. Let $X_{ij}$ denote the number of times item $i$ outperforms item $j$ out of the $K$ comparisons. Then the BT model assumes $X_{ij}\sim \text{Bin}(K,\pi_{ij})$, where 
\begin{align}\label{pij:bt}
    \pi_{ij}=\dfrac{\exp(\mu_i)}{\exp(\mu_i)+\exp(\mu_j)}
\end{align}
is the probability that $i$ is preferred over $j$, $1\leq i\neq j\leq M$. Notice that in \eqref{pij:bt}, $\bmu$ is not identifiable as $\pi_{ij}$'s remain unchanged if $\bmu=\bmu+c\bm{1}$, where $c\in\mathbb{R}$ and $\bm{1}$ is a vector with all entries $1$. Therefore, to ensure identifiability, an additional constraint is required. A common choice for identifiability constraint is $\sum_{i=1}^M \mu_i=0$ (see, \citealt{ameli2025a}).

Let $\bm{X}=(X_{ij})_{1\leq i,j\leq M}$ and $X_{ij}$'s are independent, then the likelihood function is given by
\begin{align}
\label{lh}
    \ell(\bmu\mid \mathbf{X})=\prod_{i=1}^M\prod_{j<i}\binom{K}{X_{ij}}\pi_{ij}^{X_{ij}}(1-\pi_{ij})^{K-X_{ij}}.
\end{align}
The maximum likelihood estimator (MLE) for $\bmu$ does not have a closed form and thus is obtained via iterative algorithms (\citealt{zermelo1929berechnung}, \citealt{hunter2004mm}, \citealt{newman2023efficient}).

\subsection{Bayesian BT Model} 
\label{bayesian}

In the Bayesian BT model, a prior distribution $\pi_{\bmu}(\cdot)$ for $\bmu$ is added, which may reflect the prior knowledge about the items compared. The posterior distribution  $\tilde\pi_{\bmu}(\bmu \mid \mathbf{X}) \,\propto\, \ell(\mathbf{X} \mid \bmu)\pi_{\bmu}(\bmu)$, 
quantifies the uncertainty in $\bmu$ after incorporating both the prior and the observed data $\mathbf{X}$.

The hyperparameters, the parameters of the prior distribution, can be handled by assuming a distribution for the hyperparameters (known as a hyperprior). This makes the model's structure hierarchical, enabling flexible inference. A conjugate prior provides tractable mathematical simplification; however may not be suitable for some situations. Non-conjugate priors provide greater flexibility for capturing relevant prior knowledge but may lead to a complicated posterior. In such a scenario, sampling from the posterior can be done using MCMC, for more details, see \cite{reich2019bayesian}.

\subsection{Datasets}\label{section:NFHS Dataset}
We use the National Family Health Survey-5 (NFHS-5) data, collected during 2019-21 (\citealt{NFHS5_PhaseII_Compendium_2021}). It provides information on population, nutrition, health, and family planning for Indian states/UTs. It is one of the largest household surveys of its kind in India, and is widely used to track demographic and health indicators across states/UTs of India.  

In NFHS-5, information was collected from 636,699 households, 724,115 women, and 101,839 men. The dataset is based on 131 questions (indicators) in different categories related to fertility, infant and child mortality, the practice of family planning, maternal and child health, reproductive health, nutrition, anaemia, utilisation and quality of health and family planning services. Compared to other demographic and health databases, the NFHS-5 is unique in that it not only provides individual-level records but also provides extensive information on health correlates, including socioeconomic status, access to healthcare, and risk factors.

\begin{figure}[h]
    \includegraphics[width=1\linewidth, height=9cm]{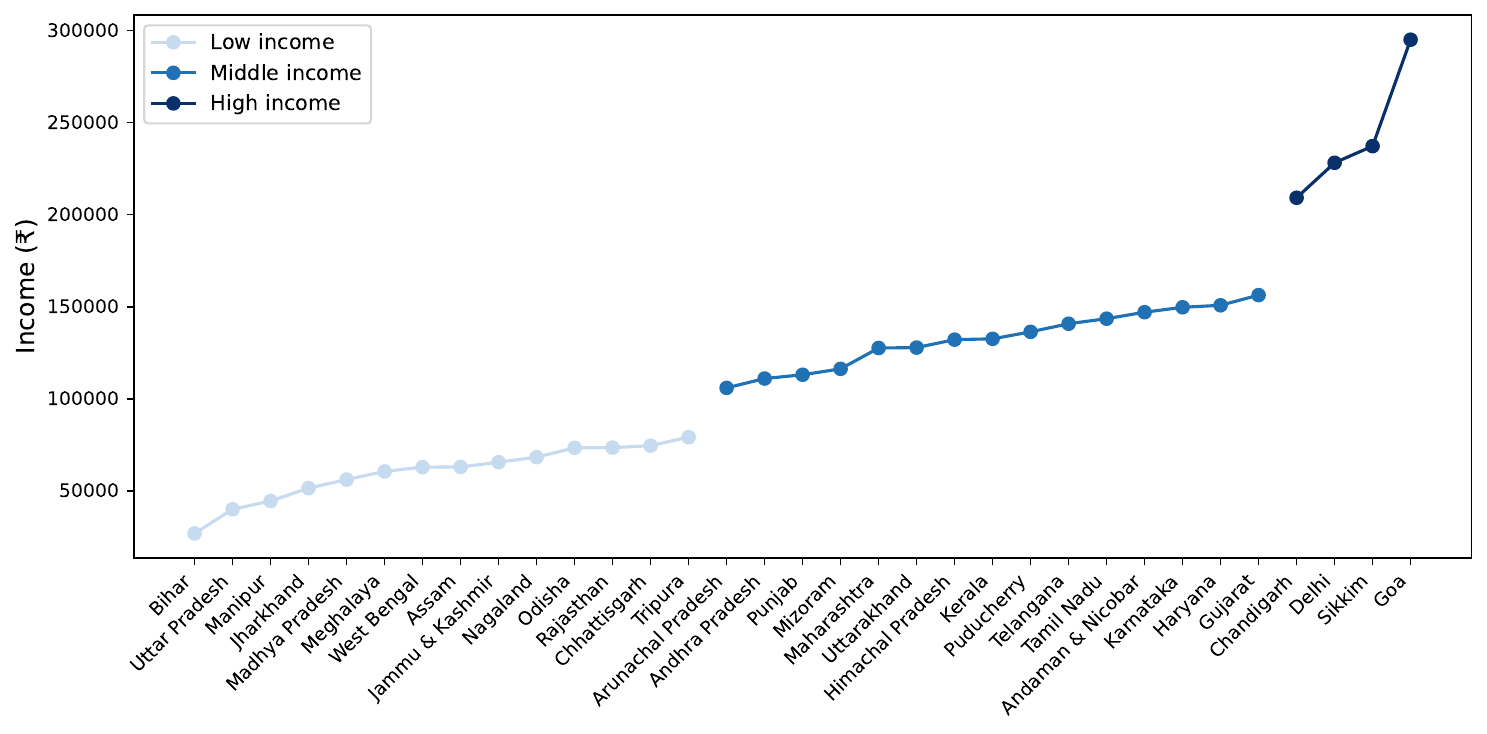}
    \caption{Per capita incomes of 33 states/UTs of India in the year 2020-2021}
    \label{per capita}
\end{figure}

The per capita income data for the states/UTs for the year 2020-2021 (same as the NFHS-5 period) are obtained from \cite{rbi_states_handbook_YEAR}. These per capita incomes are based on constant prices with a base year of 2011-12 and are presented in \Cref{per capita}. This data includes only $33$ states/UTs; the entries for Dadra and Nagar Haveli and Daman and Diu, Lakshadweep, and Ladakh are missing. We remove states/UTs with missing per capita income entries from the NFHS-5 dataset for the analysis in this paper.
\label{division}
Furthermore, \Cref{per capita} provides a natural grouping of per capita incomes across Indian states/UTs. Based on this, we refer to states/UTs ranging from Bihar to Tripura as lower-income group, those from Arunachal Pradesh to Gujarat as mid-income group, and states/UTs Chandigarh, Delhi, Sikkim, and Goa as  high-income group.

\section{Methodology}
\label{methodology_section}
In this section, we discuss the prior on merits of BT model, covariance modelling and parameter estimation by MCMC sampling methods followed by MCMC diagnostics and stopping rule.

\subsection{The Prior on Merits of BT Model} 

For different subunits of a region, the prior information may arise from spatial proximity, economic similarity, common climatic or territorial characteristics, or other historical or contextual relationships (\citealt{seymour2022bayesian}). In the context of societal development, per capita income is an important driver (see, e.g., \citealt{thorn1968per}, \citealt{hordofa2023moderating}). Therefore, in this work, we assume that the prior on $\bmu$, the merits of states/UTs, is a function of per capita income and the Gaussian distribution is a natural choice (\citealt{seymour2022bayesian}). Precisely, we assume that the prior is a zero-mean multivariate Gaussian distribution conditional on $\sum_{i=1}^M \mu_i=0$, i.e.,  
\begin{align}
\label{prior}
    (\bmu\mid \mathbf{1}^T\bmu=0)\sim \mathcal{N}_M(\bm{0}, \bm{\Omega}),\quad \text{where}\quad  \bm{\Omega}= \bm{\Sigma}-\bm{\Sigma}\mathbf{1}(\mathbf{1}^T\bm{\Sigma}\mathbf{1})^{-1}\mathbf{1}^T\bm{\Sigma},
\end{align}
and $\bm{\Sigma}$ is a function of per capita incomes (see \Cref{model_cov} for details). The distribution \eqref{prior} is degenerate, so for sampling we use a pseudoinverse (\citealt{rao1973linear}); the following section discusses the modelling of $\bm{\Sigma}$.

\subsection{Modelling \texorpdfstring{$\bm\Sigma$}{Sigma}}
\label{model_cov}
States/UTs with similar incomes are expected to have a similar growth regime (\cite{thorn1968per}). Therefore, they are expected to have similar merits, $\mu_i$'s. Consequently, in the prior, the merits of states/UTs with similar per capita incomes should have higher covariances, and the covariance should decay as the gap increases. For example, the pair Telangana and Karnataka is expected to have higher covariance than the pair Bihar and Sikkim. Kernel functions are well known for modelling covariance matrix, e.g., squared exponential, Matérn, rational quadratic (\citealt{williams2006gaussian}). We employ a squared exponential kernel to obtain $\bm{\Sigma}$. Precisely, we define the covariance between states/UTs $i$ and $j$ as
\begin{align}
\label{cov1}
    \text{cov}(\mu_i,\mu_j)&=\bm{\Sigma}_{ij}=\alpha^2\text{exp}\left(-\dfrac{d_{ij}^2}{l^2}\right)
\end{align}
where $\alpha^2$ is variance hyperparameter, $l$ is  characteristic length scale and $d_{ij}$ is a distance between $i$ and $j$. The other kernels were found to provide similar results.

Let $\rho_i$ denote the per capita income of state/UT $i$. If  $d_{ij}=|\rho_i-\rho_j|$, then the off-diagonal elements of $\bm{\Sigma}$ are very close to zero. Furthermore, $d_{ij}=|\rho_i-\rho_j|$ fails to reflect relative income disparities, i.e., the same per capita income difference between a pair of states/UTs in the low-income group and the high-income group indicates different disparity levels. For example, a  Rs. 11,000 gap is less significant between Maharashtra and Mizoram (high-income group) than between Chhattisgarh and Meghalaya (low-income group). A suitable distance function to address these issues is
\begin{align}\label{eq:dij}
d_{ij} = |\log{\rho_i} - \log{\rho_j}|.
\end{align}
For per capita incomes of Indian states/UTs, the function \eqref{eq:dij} yields off-diagonal elements of $\bm{\Sigma}$ away from zero and also captures the relative disparity levels. Therefore, we use \eqref{eq:dij} to model $\bm{\Sigma}$. Hyperparameters ($\alpha$ and $l$) can be tuned or assigned a prior distribution; in the latter case, this leads to a hierarchical model structure. In practice, numerical rounding errors can result in a covariance matrix that is not positive definite, even for valid kernel functions. This issue can be resolved through careful hyperparameter tuning.

\subsection{Estimating \texorpdfstring{$\bm\mu$}{mu} and \texorpdfstring{$\alpha^2$}{alpha}}
\label{estimation}
Due to non-conjugacy of prior \eqref{prior} on $\bmu$, the posterior does not have a closed-form expression, and direct sampling from the posterior distribution is difficult. Therefore, we employ MCMC to sample from the posterior distribution of $\bmu$ and $\alpha^2$, i.e.,
\begin{align}
\label{model}
    \tilde\pi_{\bmu,\alpha^2}(\bmu,\alpha^2\mid \mathbf{X})\;\propto\; \ell( \bmu\mid \mathbf{X})\,\pi_{\bmu}(\bmu\mid \alpha^2)\,\pi_{\alpha^2}(\alpha^2),
\end{align}
where $\ell(\cdot)$ is given in \eqref{lh}, $\pi_{\bmu}(\cdot)$ is given in \eqref{prior}, and the $\pi_{\alpha^2}(\cdot)$ is a prior on $\alpha^2$ (discussed later).

We update $\bmu$ and $\alpha^2$ in the following two steps. First, we update $\bmu$ using the Metropolis-Hastings (MH) algorithm with a modified random-walk proposal (\citealt{robert2014m}). The Gaussian prior \eqref{prior} allows us to use a proposal based on the preconditioned Crank-Nicolson (pCN) scheme (for details see Section A in the Supplement). 
Hereafter, we denote the MH algorithm with the pCN proposal by MHpCN. Specifically, MHpCN targets the marginal posterior of $\bmu$ given by
\begin{align}
\label{post:mu}
    \tilde{\pi}_{\bmu}(\bmu \mid \alpha^2,\mathbf{X})\;\propto\; \ell(\bmu\mid\mathbf{X})\,\pi_{\bmu}(\bmu\mid\alpha^2).
\end{align}
\noindent
The following result formalizes the convergence behaviour of MHpCN.
\begin{proposition}
\label{prop:pcn}
For the likelihood \eqref{lh} and prior \eqref{prior} on $\bmu$, the Markov chain $(\bmu_t)_{t \geq 0}$ generated by MHpCN has the invariant distribution \eqref{post:mu}. Furthermore, the obtained Markov chain is ergodic and convergent.
\end{proposition}

\Cref{prop:pcn} ensures that the samples generated by the MHpCN converge to the posterior distribution \eqref{post:mu} under suitable conditions on the likelihood and prior. The pCN scheme is reversible with respect to the Gaussian prior \citep{cotter2013mcmc} and has been known to be effective for high-dimensional settings \citep{hairer2014spectral}. These properties contribute to the efficiency of MHpCN observed in various applications (\citealt{pezzetti2025function}).

Next, for estimating $\alpha^2$, we take an inverse-gamma prior on $\alpha^2$, which is conjugate to the Gaussian distribution. Thus, we get a closed-form expression for the conditional posterior given  $\bmu$. Consequently, $\alpha^2$ can be estimated using Gibbs sampling. The conditional posterior distribution of $\alpha^2$ is given by
\begin{align}
\label{invgamma}
    (\alpha^2\mid \bmu)\sim \text{Inv-Gamma}(\chi+M/2,\omega+\bmu^T \bm{\Omega}^{-1}\bmu),
\end{align}
where $\chi$ and $\omega$ are shape and scale parameters, respectively, for the inverse gamma distribution, and $\bm{\Omega}$ is given in \eqref{prior}. Therefore, we use a Metropolis-within-Gibbs sampling scheme, where $\bmu$ is updated using MHpCN and $\alpha^2$ is updated via Gibbs sampling. The complete algorithm is provided in \Cref{alg:mcmc}. Natural estimators of $\bmu$ and $\alpha^2$ are the corresponding sample means obtained via \Cref{alg:mcmc}.

\begin{algorithm}
\caption{MCMC for model parameters}
\label{alg:mcmc}
\begin{algorithmic}[1]
\Require Initial model parameters $\bmu^{(0)}$ and $\alpha^2_{(0)}$, tuning parameter $\beta \in (0,1)$,
data $\mathbf{X}$ through win matrix, $\bm\Omega$, hyperparameters $(\omega,\chi).$

\For{$t^{th}$ iteration}

    \State \textbf{(Gibbs step)} Sample variance parameter $\alpha^2_{(t)}$ from its inverse-gamma full conditional conditioned on $\bmu^{(t-1)}$.

    \State \textbf{(pCN proposal)}  
    Propose
    \[
    \mathbf{\bmu}^\ast
    =
    \sqrt{1-\beta^2}\,\mathbf{\bmu}^{(t-1)}
    +
    \beta\,\bm\xi,
    \quad
    \bm{\xi} \sim \mathcal{N}(\mathbf{0},\alpha^2_{(t)}\bm{\Omega}).
    \]

    \State \textbf{(Likelihood evaluation)}  
    Compute the log-likelihood at $\bmu^\ast$.

    \State \textbf{(MH step)}  
    Accept $\bmu^\ast$ with probability
    \[
    \min\!\left(1, \log{
    \ell( \bmu^*\mid \mathbf{X})}-\log{
         \ell(\bmu^{(t-1)}\mid \mathbf{X}})
    \right).
    \]

\EndFor \,(with a stopping criteria).

\State \Return posterior samples $\{\bmu^{(t)}\}$ and $\{\alpha^2_{(t)}\}$.

\end{algorithmic}
\end{algorithm}

Let $\{\bmu_t\}_{t=1}^\infty$ denotes the Markov chain obtained from \Cref{alg:mcmc} corresponding to $\bmu$. Denote the mean of $\bmu_t$ by $\mathbb{E}_{\pi}[\bmu_t]$ and define $\bm{\Sigma}_{\bmu}=\text{Cov}(\bmu_t)$. A natural estimator for $\bmu$ is 
$$\bar{\bmu}_N = \frac{1}{N}\sum_{t=1}^N \bmu_t. $$
The asymptotic properties of $\bar{\bmu}_N$ are well documented, see \cite{geyer2011introduction}.  If the chain, $\{\bmu_t\}_{t=1}^\infty$, satisfy Harris
ergodicity with invariant distribution $\tilde\pi_{\bmu}$ and 
$\mathbb{E}_{\pi}[\|\bmu_t\|^2] < \infty$,
then $\bar\bmu_N \to \mathbb{E}_{\pi}[\bmu_t]$ almost surely as $N\to\infty$. 
Furthermore, if there exists $\delta>0$ such that
$\mathbb{E}_{\pi}[\|\bmu_t\|^{2+\delta}] < \infty$
and $\{\bmu_t\}_{t=1}^\infty$ is polynomially ergodic of order $m>(2+\delta)/\delta$, then
\[
\sqrt{N}\bigl(\bar\bmu_N - \mathbb{E}_{\pi}[\bmu_t]\bigr)
\xrightarrow{d} \mathcal{N}(\bm{0}, \bm{L}),
\]
where 
\begin{align}
\label{auto}
    \bm{L}=\bm{\Sigma}_{\bmu}+\sum_{k=1}^{\infty}(L_k+L_k^T), \quad \text{and} \quad L_k=\text{Cov}(\bmu_t,\bmu_{t+k}).
\end{align}
For more details on properties of $\bar\bmu_N$, please see \cite{vats2019multivariate} and references therein.

\subsection{MCMC diagnostics}
\label{diagnostics}
\Cref{prop:pcn} guarantees the asymptotic convergence of the MHpCN algorithm. In practice, however, the algorithm is executed for a finite number of iterations, and estimates are approximated using sample averages. This inevitably introduces some error, and quantifying this error is therefore crucial for assessing the accuracy and reliability of the resulting estimates.


To compare the dependent Markov chain $\{\bmu_t\}_{t=1}^N$ and an i.i.d. sample, the effective sample size (ESS) is used as a measure of sampling efficiency \cite{vats2019multivariate}. The ESS is the number of i.i.d. observations required to achieve the same variance as the sample mean from the Markov chain. Note that for i.i.d samples, $L=\Sigma_{\bmu}$ in \eqref{auto}. For full-rank $\bm{\Sigma}_{\bmu}$ and $\bm{L}$, the ESS is defined  as
\begin{align*}
   \text{ESS}=N\left[\dfrac{\text{det}(\bm{\Sigma}_{\bmu})}{\text{det}(\bm{L})}\right]^{1/M}
\end{align*}
where $\det(\cdot)$ denotes the determinant of a matrix. 
For singular $\bm{\Sigma}_{\bmu}$ and $\bm{L}$, \cite{mukherjee_mvessz} provided an alternate approach for ESS. Consider the spectral decomposition $\bm{\Sigma}_{\bmu}=\bm{U}\bm{D}\bm{U}^T$ where $\bm{D}=\text{diag}(e_1,\ldots,e_r,0,\ldots,0)$, $e_i$ are the positive eigenvalues of $\bm{\Sigma}_{\bmu}$ and $\bm{U}=(\bm{u}_1, \ldots,\bm{u}_r,\ldots)$, a matrix whose $i^{th}$ column is orthonormal eigenvectors of $\bm{\Sigma}_{\bmu}$ corresponding to eigenvalue $e_i$.
Let $\bm{U}_r$ is a $M\times r$ matrix with $\bm{u}_i$ as $i^{th}$ column, $\bm{\Sigma}_{\bmu}^S= \bm{U}_r^T\bm{\Sigma}_{\bmu}\bm{U}_r$ and 
$\bm{L}_S= \bm{U}_r^T\bm{L}\bm{U}_r$. Then, \cite{mukherjee_mvessz} defined the multivariate ESS as
\begin{align}
\label{mvess}
    \text{ESS}= N\left[\dfrac{\text{det}(\bm{\Sigma}_{\bmu}^S)}{\text{det}(\bm{L}_S)}\right]^{1/r}.
\end{align}
To estimate the ESS, we replace $\bm\Sigma_{\bmu}$ and $\bm L$ with their corresponding estimators. \cite{vats2018strong} provides following estimators of $\bm{\Sigma}_{\bmu}$ and $\bm{L}$,
$$\hat{\bm{\Sigma}}_{\bmu}=\dfrac{1}{N-1}\sum_{t=1}^N (\bmu_t-\bar{\bmu}_N)(\bmu_t-\bar{\bmu}_N)^T,\quad \text{and} \quad 
\hat L=\hat{\bm{\Sigma}}_{\bmu}+\sum_{k=1}^{b-1}w_k (\hat{L}_k+\hat{L}_k^T),
$$
where $\hat{L}_k=(n - k)^{-1} \sum_{t=1}^{n-k} (\bmu_t - \bar{\bmu}_N)(\bmu_{t+k} - \bar{\bmu}_N)^{\top}$ for $k\geq 1$ and $b$ is the batchsize after which there is no significant autocorrelation in the chain and $w_k$ is the lag window that assigns weight to the lags. \cite{vats2018strong} took $b=b_N$ such that $b_N\rightarrow\infty$ and $b/N\rightarrow0$ and  showed that the estimators converge almost surely.

In practice, the eigenvalues may be very close to $0$ but not exactly equal to $0$. Therefore, the rank $r$ is estimated by thresholding the eigenvalues, and $\bm{U}_{\hat{r}}$ can be obtained by retaining the eigenvectors whose corresponding eigenvalues are greater than the threshold \citep{mukherjee_mvessz}. Consequently, we have $\hat{\bm{\Sigma}}_{\bmu}$, $\hat{\bm{L}}$,  $\hat{\bm{\Sigma}}_{\bmu}^{\hat{S}}=\bm{U}_{\hat{r}}\hat{\bm{\Sigma}}_{\bmu}U_{\hat{r}}$ and $\hat{\bm{L}}_{\hat{S}}=\bm{U}_{\hat{r}}\hat{\bm{L}}\bm{U}_{\hat{r}}$ where $\hat{r}$ is the estimated rank. Thus the ESS is estimated using \eqref{mvess} by
\begin{align}
\label{mvessz}
    \hat{\text{ESS}}= N\left[\dfrac{\text{det}(\hat{\bm{\Sigma}}_{\bmu}^{\hat{S}})}{\text{det}(\hat{\bm{L}}_{\hat{S}})}\right]^{1/\hat{r}}.
\end{align}

Another diagnostic tool for MCMC is the acceptance ratio (\citealt{roberts2001optimal}). For \Cref{alg:mcmc}, the acceptance ratio reduces to the ratio of likelihoods, and for multivariate chains, its desirable value lies between 20\% and 30\% (\cite{cotter2013mcmc}). A very high acceptance rate often indicates that the chain is stuck in a specific region and is not mixing properly. Furthermore, a very low acceptance rate indicates that the chain is far from stationarity. For better results, the Effective Sample Size (ESS) should be sufficiently large. 

\subsection{Stopping Rule}
In addition to standard MCMC diagnostics, a practical implementation necessitates an appropriate stopping rule, as standard criteria may not be well-suited to the problem at hand. We therefore propose a problem-specific stopping criterion based on the stability of the induced state rankings. The Kendall tau distance (\citealt{cicirello2020kendall}) is used to define the stopping criteria. Let $\tau_t$ denote the ranking induced by $\bar\bmu_t$ at iteration $t$, and $K_d(\tau_t, \tau_{u})$ denote the Kendall tau distance between rankings $\tau_t$ and $\tau_u$. 
\begin{definition}[$m$-stable ranking]
Let the \Cref{alg:mcmc} be run for $N$ iterations.
The ranking is said to be $m$-stable if
\[ K_d(\tau_t, \tau_{t+1}) = 0 ,\]
for $t=N-m,N-m+1,\ldots, N-1$.
\end{definition}
If the ranking is $m$-stable for large $m$, we terminate the algorithm. In applications, we take $m$ as an increasing function of $N$. Finally, we note that the merit estimates may continue to exhibit variability even after the $m$-stable ranking is achieved for large $m$.

\section{Analyzing NFHS-5 data}
\label{NFHS_implementation}
The dataset comprises $131$ indicators for each state and union territory. There are $6$ indicators with $5$ or more missing entries; they were excluded from the study. Additionally, entries for 9 more indicators were missing for Chandigarh, so we conducted the analysis by removing these indicators to include Chandigarh. Further, we also analysed the data by removing Chandigarh and removing only 6 indicators mentioned earlier. 

For each indicator, pairwise comparisons are constructed between states/UTs, with the winner determined by the direction of desirability of the indicator. For instance, higher values correspond to better performance for vaccination coverage, whereas lower values are preferred for diarrhoea prevalence. Aggregating these comparisons across all indicators yields the win matrix. The covariances \eqref{cov1} for the prior are constructed using the squared exponential kernel. The \Cref{alg:mcmc} is run for $3$ million iterations, and uses $m=1.9$ million for the stopping criteria. We use Python for the implementation.

Section \ref{ranking:all} presents the overall rankings of states/UTs using both Bayesian BT and BT models, along with a secondary analysis excluding Chandigarh. To assess performance within comparable economic contexts, states and UTs are grouped by income levels as defined in \Cref{division}, and ranked accordingly in \Cref{ranking:subset:states/UTs}. This intra-group analysis highlights shifts in relative positions that arise when comparisons are restricted to similar financial capacities. A case study of Gujrat vs. Karnataka is presented in \Cref{case_study} followed by simulation study \Cref{simulation} to assess the prior's sensitivity.

\subsection{Ranking of all states/UTs}
\label{ranking:all}
\begin{figure}[h]
    \centering
\includegraphics[width=0.5\linewidth,height=6cm]{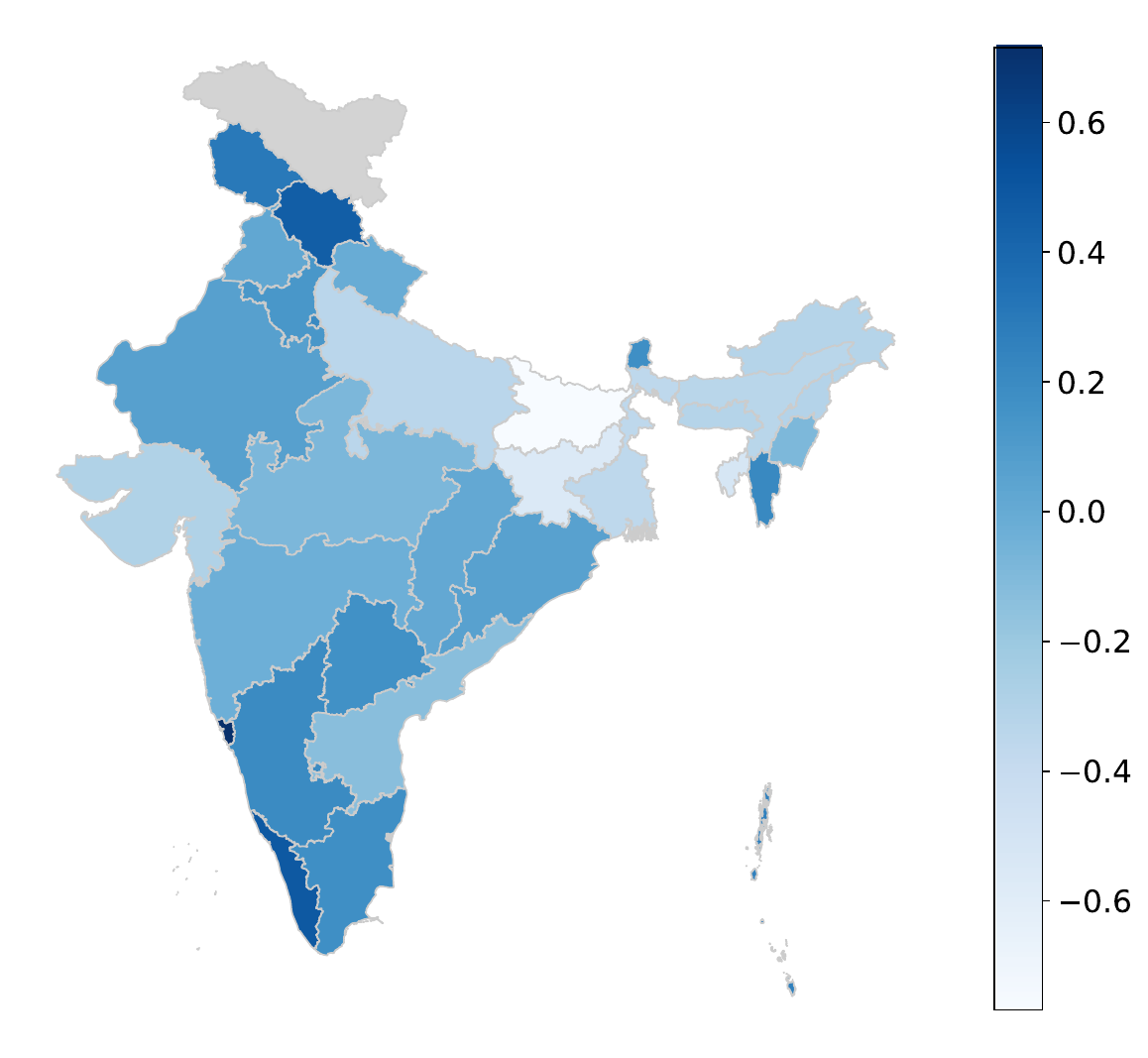}
        \caption{Heatmap of estimated merit parameter values}
        \label{image:heatmap}
\end{figure}

First, we analyse by keeping Chandigarh in the dataset. Here, we have 116 indicators to construct the win matrix. For implementation, we set $ l=0.09$, $\chi=2$, $\omega=1$, and $\beta=0.009$. Guided by traceplots (See Section D in the Supplement), the first $1$ million iterations are discarded as burn-in. More details on implementation are provided in Section C of the Supplement. The estimated merits are displayed as a heatmap in \Cref{image:heatmap}, illustrating the relative performance of the states/UTs.
\begin{figure}[h]
    \centering
    \includegraphics[width=1\linewidth,height=8cm]{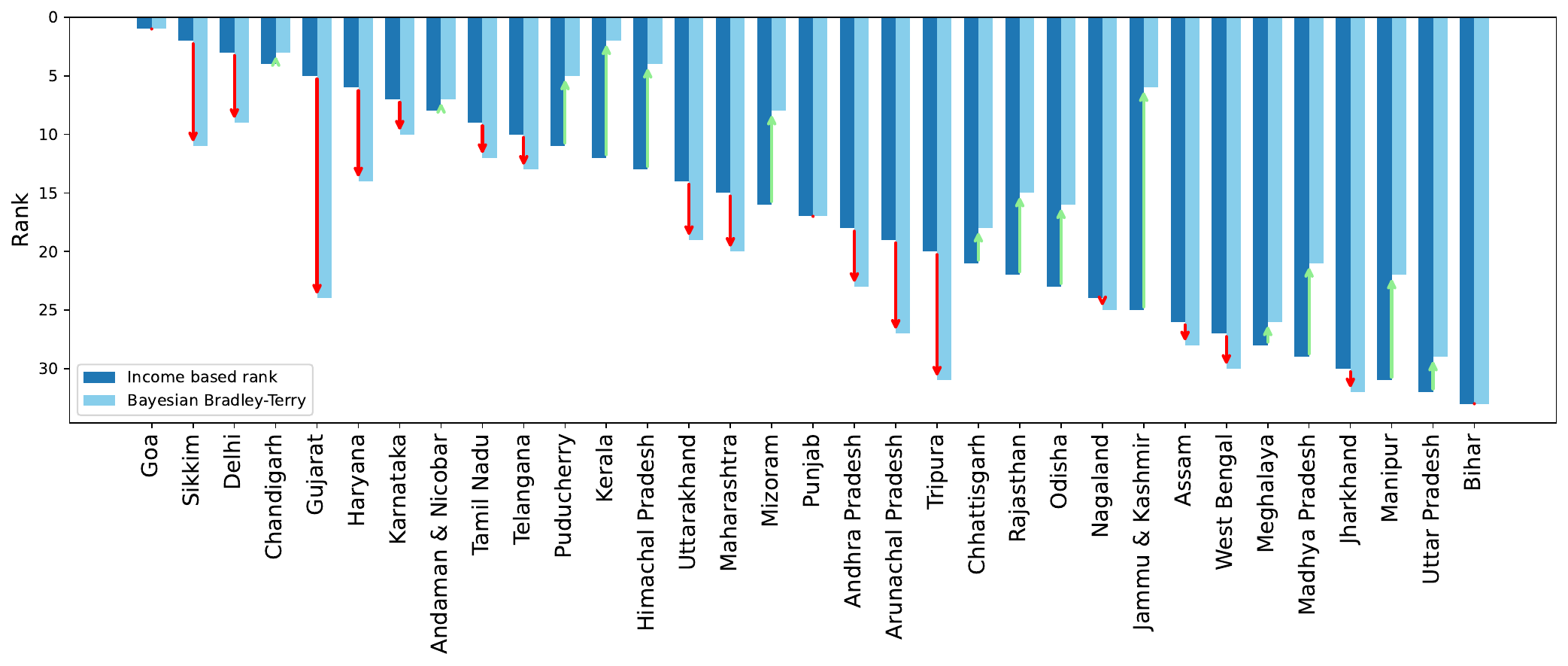}
    \caption{Rankings obtained via per-capita incomes and Bayesian BT model}
    \label{rankplots_overall}
\end{figure}
\noindent

\begin{figure}[H]
    \centering
\includegraphics[width=1\linewidth,height=7.5cm]{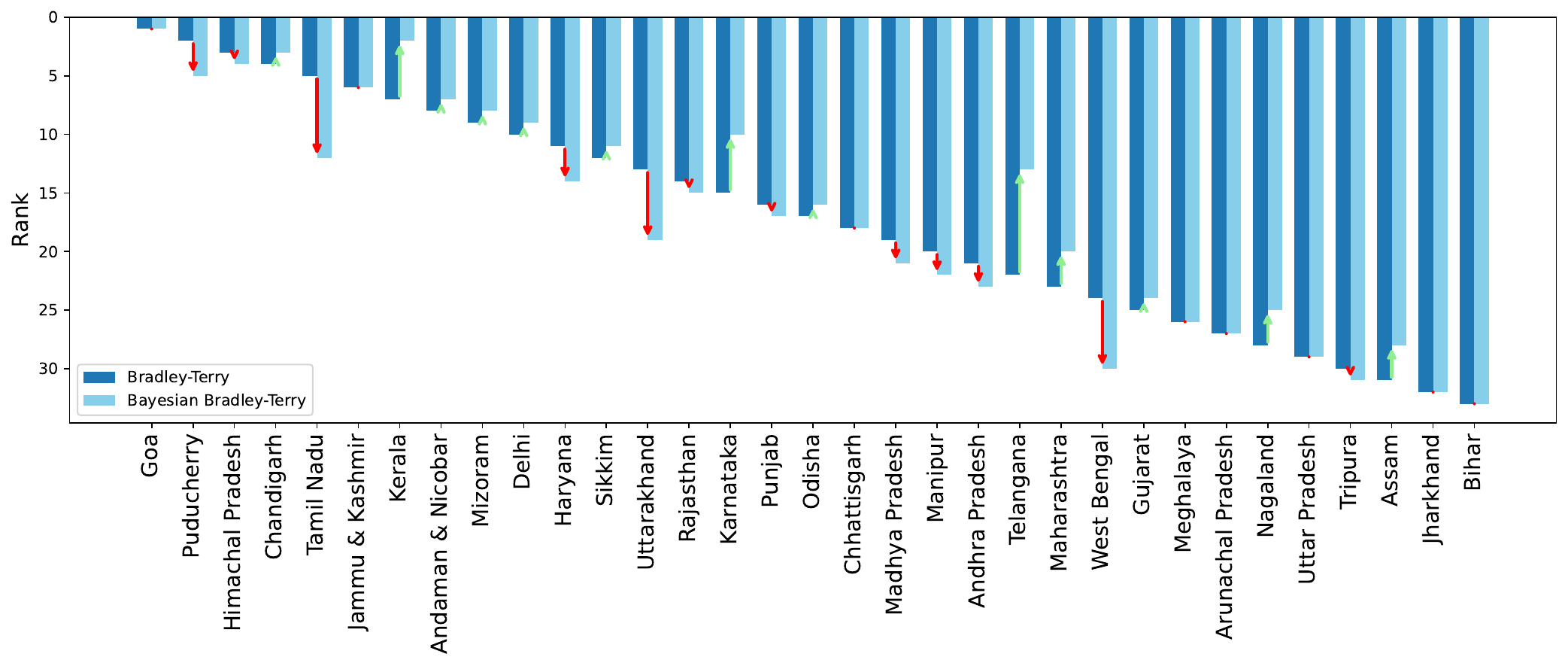}
    \caption{Rankings obtained via BT and Bayesian BT models}
    \label{bt-bbt}
\end{figure}

Further, we implemented the BT model on the same set of indicators without the prior distribution. The BT model parameters are estimated using the scheme proposed by \cite{newman2023efficient}. A comparison between the Bayesian BT and BT rankings is provided in \Cref{bt-bbt}. A closer look at \Cref{bt-bbt} reveals several interesting insights. The states/UTs at the extremes of the rankings remain consistent across both approaches, while minor reorderings occur in the middle. For example, Telangana's rank improves under the Bayesian BT model, illustrating the influence of the prior. This indicates that while the Bayesian ranking is driven by the likelihood, it is also significantly informed by the prior.

Excluding Chandigarh allows the inclusion of 9 additional indicators, bringing the total indicator count to 125. The resultant ranking is provided in Figure 7 of the Supplement. Comparing them, we observe no significant differences between the two rankings, suggesting that the remaining 9 indicators provide only limited information.

\subsection{Ranking of subsets of states/UTs}
\label{ranking:subset:states/UTs}
We separately consider subsets of states/UTs corresponding to low and middle income groups, as well as the set excluding very high-income states/UTs (Goa, Delhi, Sikkim, and Chandigarh), as defined in \Cref{section:NFHS Dataset}. The rankings are obtained using the same methodology as in \Cref{ranking:all}, and the results including those in the previous section are presented in \Cref{comparison}.

\begin{table}[h]
\scriptsize
\caption{Rankings of States/UTs under different models}
\label{comparison} 
\begin{tabular}{|l|c|c|c|c|c|c|}
\toprule
\makecell{States/UTs} 
& \makecell{Per capita\\ Income\\ Rank} 
& \makecell{Overall\\ ranking\\ (33 states)} 
& \makecell{Low level\\ income states\\ (14 states)} 
& \makecell{Middle level\\ income states\\ (15 states)} 
& \makecell{States without\\ high incomes\\ (29 states)} 
& \makecell{BT model} \\
\midrule
Andaman \& Nicobar & 8  & 7  & -- & 6  & 5  & 8  \\
Andhra Pradesh     & 18 & 23 & -- & 13 & 19 & 22 \\
Arunachal Pradesh  & 19 & 27 & -- & 15 & 21 & 27 \\
Assam              & 26 & 28 & 8  & -- & 25 & 31 \\
Bihar              & 33 & 33 & 14 & -- & 29 & 33 \\
Chandigarh         & 4  & 3  & -- & -- & -- & 4  \\
Chhattisgarh       & 21 & 18 & 2  & -- & 14 & 18 \\
Goa                & 1  & 1  & -- & -- & -- & 1  \\
Gujarat            & 5  & 24 & -- & 14 & 20 & 25 \\
Haryana            & 6  & 14 & -- & 9  & 10 & 11 \\
Himachal Pradesh   & 13 & 4  & -- & 3  & 2  & 3  \\
Jammu \& Kashmir   & 25 & 6  & 1  & -- & 4  & 6  \\
Jharkhand          & 30 & 32 & 13 & -- & 28 & 32 \\
Karnataka          & 7  & 10 & -- & 7  & 6  & 15 \\
Kerala             & 12 & 2  & -- & 1  & 1  & 7  \\
Madhya Pradesh     & 29 & 21 & 5  & -- & 17 & 20 \\
Maharashtra        & 15 & 20 & -- & 12 & 16 & 19 \\
Manipur            & 31 & 22 & 6  & -- & 18 & 21 \\
Meghalaya          & 28 & 26 & 9  & -- & 23 & 26 \\
Mizoram            & 16 & 8  & -- & 5  & 7  & 9  \\
Nagaland           & 24 & 25 & 11 & -- & 22 & 28 \\
Delhi              & 3  & 9  & -- & -- & -- & 10 \\
Odisha             & 23 & 16 & 4  & -- & 12 & 17 \\
Puducherry         & 11 & 5  & -- & 2  & 3  & 2  \\
Punjab             & 17 & 17 & -- & 10 & 13 & 16 \\
Rajasthan          & 22 & 15 & 3  & -- & 11 & 14 \\
Sikkim             & 2  & 11 & -- & -- & -- & 12 \\
Tamil Nadu         & 9  & 12 & -- & 8  & 8  & 5  \\
Telangana          & 10 & 13 & -- & 4  & 9  & 23 \\
Tripura            & 20 & 31 & 12 & -- & 27 & 30 \\
Uttar Pradesh      & 32 & 29 & 7  & -- & 24 & 29 \\
Uttarakhand        & 14 & 19 & -- & 11 & 15 & 13 \\
West Bengal        & 27 & 30 & 10 & -- & 26 & 24 \\
\bottomrule 
\end{tabular}

\end{table}

A comparison of the subset ranking for low-income states/UTs with the overall ranking (\Cref{rankplots_overall}) reveals some notable positional shifts, particularly between Uttar Pradesh and Nagaland. In the overall ranking, Nagaland is ranked below Uttar Pradesh; however, when the analysis is restricted to low-income states/UTs, Nagaland attains a higher rank. This shift is due to the fact that Nagaland performs relatively better than Uttar Pradesh when compared against high-income states/UTs (e.g., Chandigarh, Goa, Gujarat), whereas Uttar Pradesh exhibits stronger performance primarily within the group of lower-income states/UTs. Moreover, the substantial income disparity between these states ensures that their ranks remain non-consecutive across both rankings.

The subset ranking among middle-income states/UTs also exhibits some differences relative to the overall ranking. In particular, Telangana shows a marked improvement, moving up by $4$ positions when the analysis is restricted to middle-income states. Although its performance on NFHS indicators remains moderate compared to states such as Mizoram, Tamil Nadu, Karnataka, and Kerala, the strong correlation structure within this income group leads to increased influence of the prior distribution on the estimates. As a result, Telangana, having an income level comparable to Kerala, Tamil Nadu, Andaman \& Nicobar, and Karnataka, appears adjacent to these states in the ranking. However, when the covariance is reduced by decreasing the length-scale parameter, this dependence weakens, and Telangana shifts away from these states, aligning more closely with its position in the overall ranking.

We repeat the analysis after excluding states/UTs with high incomes (Goa, Delhi, Sikkim, and Chandigarh). The relative positions of most states remain unchanged, with only minor swaps observed among adjacent pairs such as Assam and Uttar Pradesh, Meghalaya and Arunachal Pradesh, and Mizoram and Karnataka. These changes likely reflect small perturbations in posterior estimates following the exclusion. Importantly, no substantial rank reversals are observed, indicating that the overall model structure is robust and not driven by high-income outliers.

\begin{remark}
Minor swaps are observed in the subset rankings, for example, between Meghalaya and Assam among lower-income states, arising from changes in the likelihood when comparisons with high income states are excluded. Since the differences among some of the estimated merit parameters are small, the ranking based on posterior means is sensitive to slight shifts in the posterior distribution. 
\end{remark}

\subsection{A case study: Gujarat vs. Karnataka}
\label{case_study}
Gujarat, despite a high per capita income of Rs. 156,285, ranks 25th, whereas Karnataka, with a comparable income level, ranks 10th. The relatively poor overall rank of Gujarat, despite being a high-income state, has also been noted in several studies, e.g., \cite{viswanathan2021growth} analyzed the mismatch between economic growth and human development outcomes,  \cite{iyengar2016determines} analyzed health facilities. In these studies, Gujarat is compared with other states, such as Kerala and Tamil Nadu, the states' income was not taken into account. We compare Gujarat's performance with that of Karnataka, which has a comparable per capita income.

A comparison of NFHS indicators shows that Gujarat underperforms Karnataka in several dimensions, including women’s literacy, empowerment, higher educational attainment, and internet usage, highlighting gaps in educational access and gender equity. It also lags in public health indicators such as disease prevalence (e.g., anemia), cancer screening, and access to public healthcare, suggesting the need for strengthened healthcare infrastructure and outreach. More details about the indicators are provided in Section E of the Supplement.

\subsection{Simulation study}
\label{simulation}
To assess sensitivity to the prior, we conduct a simulation study. The number of comparisons per pair is fixed at $K=130$, so that $W_{ij} \sim \text{Bin}(130, \pi_{ij})$ (see \Cref{BTM}). True merit parameters are generated from the prior distribution using different lengthscales $l \in \{0.007,0.05,0.09,0.9,1.5,3\}$. These parameters are used to simulate the win matrix, which is subsequently used to estimate merits using the proposed Bayesian BT model (with $l=0.09$) and the standard BT model.

To capture deviation of estimated $\bmu$ from the true model, we define
$$\delta= \max_{1\leq i\leq 33}|\mu_i^{true}-\mu_i^{estimated}| $$
Figure 30 in the Supplement presents $\delta$ for the BT and Bayesian BT models at different values of the true $l$. We observe that for small $l$, the Bayesian BT model exhibits a larger $\delta$ than the BT model, reflecting a mismatch between the assumed dependence structure and the near-independence in the data. As the true $l$ increases, the error of the Bayesian BT model decreases fast and eventually falls below that of the BT model. This behaviour arises because larger $l$ induce similarity among states with comparable incomes, aligning with the prior specification. Therefore, the Bayesian BT model is sensitive to the choice of lengthscale $l$. 


\section{Discussion}
\label{discussion}
To the best of our knowledge, no sound ranking methodology based on a large number of indicator variables is available in the existing literature. To address this, we develop a methodology using the Bayesian Bradley–Terry (BT) framework. 
The Bayesian BT framework has been used to rank entities in different fields, e.g., comparisons of machine learning algorithms (\citealt{wainer2023bayesian}) and comparisons of deprivation among urban regions (\citealt{seymour2022bayesian}). However, how to incorporate multiple indicators in the BT framework remained an open problem. Our study addresses this problem by converting indicators to paired comparisons. Our estimation methodology uses an MCMC approach based on a preconditioned Crank–Nicolson algorithm and Gibbs sampling.  The proposed estimator is consistent. Furthermore, we propose a problem-specific stopping rule based on ranking stability. 

We apply the proposed methodology to NFHS-5 data to rank states/UTs of India using multiple development indicators (from NFHS-5). The prior covariance is specified as a function of per capita income, capturing the notion that states/UTs with similar income levels exhibit higher covariance, with covariance decaying as income differences increase. The methodology is applied to the full dataset and to subsets of states/UTs within similar income groups to enable a more detailed comparison.  
In the current study, prior covariance is a function of per capita income only; however, the model can be easily extended to include several covariates to define the prior distribution.

The Bayesian BT and BT rankings exhibit notable variation across states, offering important comparative insights. In particular, economic standing does not necessarily translate into a high rank; for instance, Gujarat, despite being a high-income state, ranks substantially lower, indicating that the methodology captures dimensions beyond income. 
Consequently, the proposed framework offers several policy-relevant insights. For instance, Gujarat's lower rank is due to weaker outcomes in health and education indicators, highlighting areas requiring targeted intervention. More generally, the rankings can help identify sector-specific deficiencies, such as gaps in healthcare access, education, or gender outcomes, even among economically advanced regions. Subgroup analyses by income level further enable equitable benchmarking by comparing states with similar resource levels.
Overall, the framework provides a data-driven basis for targeted policy design, efficient resource allocation, and monitoring of regional development.

The proposed ranking methodology is readily applicable to several other domains, e.g., the public-health and education. By enabling coherent comparisons across multiple indicators, it provides actionable insights for targeted policy interventions and efficient resource allocation. Finally, there are several interesting directions for extending this study; some are discussed below.
\begin{enumerate}
\item The proposed methodology assigns equal weight to all indicators in constructing the win matrix; however, this assumption may be restrictive in practice. Indicators differ in importance, e.g., those related to education may warrant greater emphasis than those capturing tobacco or alcohol use. Incorporating indicator-specific weights offers a more realistic and flexible ranking framework. 

\item Several indicators capture overlapping information (e.g., infant mortality and related child health measures), whereas the current framework assumes independence among $X_{ij}$. Such redundancy can implicitly overweight certain sectors and introduce bias. Restricting attention to indicators with largely independent information, or explicitly accounting for their dependence, is likely to yield more balanced and parsimonious rankings.

\item The NFHS-5 data exhibit intransitivities in pairwise comparisons; for example, in terms of number of indicators of superiority, Andhra Pradesh outperforms Chhattisgarh, Chhattisgarh outperforms Madhya Pradesh, yet Madhya Pradesh outperforms Andhra Pradesh. The BT model does not account for such inconsistencies, which may influence the resulting rankings. Extending the framework to explicitly model intransitivity presents a natural direction for further methodological development, see, e.g., \cite{spearing2023modeling}.
\end{enumerate}

\section*{Conflicts of Interest}
No Conflict of interest to declare.

\section*{Data availability}
The NFHS-5 data used in this article are publicly available on
Open Government Data (OGD) Platform India
\href{https://www.data.gov.in/catalog/national-family-health-survey-nfhs-5-provisional}
{(https://www.data.gov.in/catalog/national-family-health-survey-nfhs-5-provisional)}. The per-capita income data are publicly available at Niti Aayog website (\href{https://iced.niti.gov.in/economy-and-demography/key-economic-indicators/socio-economic}{https://iced.niti.gov.in/economy-and-demography/key-economic-indicators/socio-economic}).

\section*{Supplementary material}
Supplementary material includes the proof of Proposition \ref{prop:pcn}, and some additional details of NFHS-5 data analysis.

\section*{Acknowledgments}
 The work of Rahul Singh was supported by New Faculty Seed Grant No. MI03038G at Indian Institute of Technology Delhi, India. 

\section*{Declaration of AI use}
The authors used {\it Gemini} in order to improve the readability and edit the language of the manuscript. After using this tool, the authors reviewed and edited the content as needed and take full responsibility for the article's content.


\bibliographystyle{apalike}
\bibliography{ref}

\includepdf[pages=1-20]{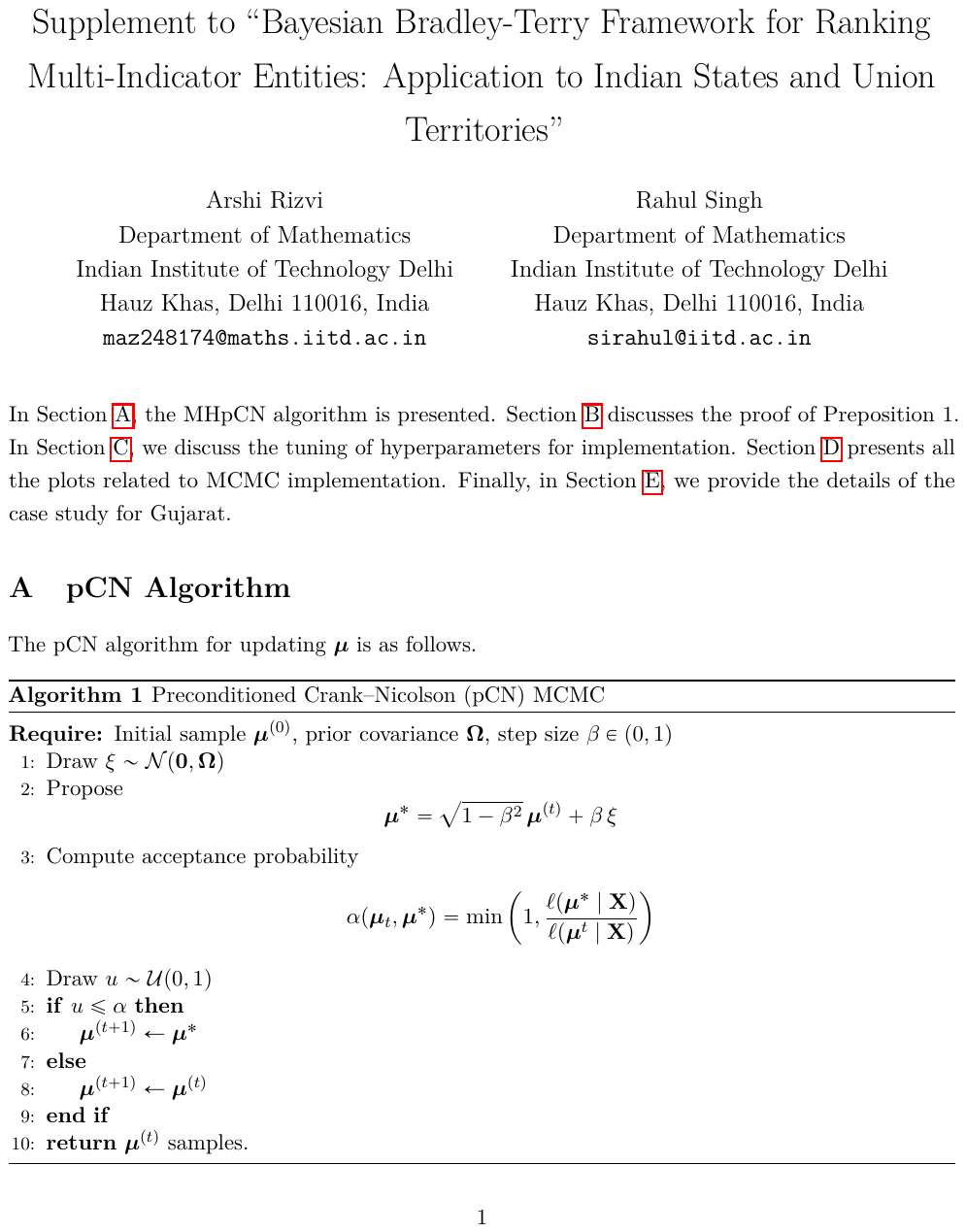}

\end{document}